\documentclass[aps,prd,showpacs,preprint,groupedaddress]{revtex4}
\usepackage{amssymb,amsmath}
\usepackage{graphicx}

\newcommand{\bra}[1]{\langle #1|}
\newcommand{\ket}[1]{|#1\rangle}
\newcommand{\expect}[1]{\langle #1 \rangle}

\begin{document}


\title{Inflaton decay in an alpha vacuum}

\author{S. Naidu}
\email{svn@andrew.cmu.edu}
\author{R. Holman}
\email{rh4a@andrew.cmu.edu}
\affiliation{Department of Physics, Carnegie Mellon University, Pittsburgh PA 15213}
\date{\today}

\begin{abstract}
We study the alpha vacua of de Sitter space by considering the decay
rate of the inflaton field coupled to a scalar field placed in an
alpha vacuum. We find an {\em alpha dependent} Bose enhancement
relative to the Bunch-Davies vacuum and, surprisingly, no
non-renormalizable divergences. We also consider a modified alpha
dependent time ordering prescription for the Feynman propagator and
show that it leads to an alpha independent result.  This result
suggests that it may be possible to calculate in any alpha vacuum if
we employ the appropriate causality preserving prescription.
\end{abstract}


\maketitle

\section{Introduction}

Field theory in de Sitter space is an area of active interest as the
inflationary epoch can be well approximated by de Sitter
space. Further, de Sitter space may even be applicable to the current
universe as it is known that the expansion is
accelarating\cite{Riess:1998cb,Perlmutter:1998np}. Current models of
inflation postulate a scalar field rolling to the bottom of its
potential as the source of the energy density that drives the
expansion and quintessence models attempt to do the same for the
present accelaration observed in the universe. Further, in inflation
it is the quantum fluctuations in the scalar field that seeds the
perturbations responsible for the temperature anisotropy and large
scale structure and observations are in agreement with the predictions
of inflation\cite{Peiris:2003ff}. The nature of the de Sitter
invariant vacuum is tied to trans-Plankian effects\cite{Martin:2003kp}
in inflation. For these purposes it is important to investigate the
formal construction of field theory in de Sitter space.

Field theory in de Sitter space and the question of possible vacua of
free scalar field theory in de Sitter space have been investigated by
\cite{Chernikov:1968zm, Tagirov:1973vv, Mottola:1985ar, Allen:1985ux}.
One of the surprising results that emerged was a vacuum choice
ambiguity that is absent in Minkowski space. They found that in de
Sitter space there exists an entire family of de Sitter invariant
vacua that can be used to construct a consistent free field
theory. This vacuum ambiguity is discussed for general spacetimes by
Long and Shore \cite{Long:1996wf} using a wavefunctional formalism. If
one insists that the vacuum reduce to the Minkowski vacuum in the
limit that the expansion vanishes then a single member of the family
is selected and is commonly denoted the Euclidean or Bunch-Davies
vacuum \cite{Bunch:1978yq}. The vacuum choice ambiguity is normally
addressed in inflationary calculations by assuming that the vacuum is
the distinguished Bunch-Davies vacuum and this leads to successful
predictions of the temperature anisotropies. The question of what
physics is responsible for the vacuum selection remains to be
addressed and a number of papers address the use of the general vacuum
with a cut-off to encode trans-Plankian effects\cite{Easther:2002xe,
Danielsson:2002kx, Goldstein:2002fc, Kaloper:2002uj}. Interacting
theories in the general vacua, however, are argued \cite{
Einhorn:2002nu, Banks:2002nv, Collins:2003zv} to be inconsistent
usually due to the presence of intractable divergences.

Generally the dynamics of inflation are studied treating the inflaton,
the scalar field driving inflation, as a classical homogenous field
and the gaussian approximation is applied to the quantum fluctuations
about this configuration. As mentioned these fluctuations seed the
perturbations in the universe. For a review of the quantum treatment
of inflationary dynamics see \cite{Boyanovsky:1997yh}. Of particular
interest to this paper is the question of particle decay in de Sitter
space and \cite{Boyanovsky:1997ab} investigates this questions
assuming that the Bunch-Davies vacuum is the relevant one. The same
problem is addressed in a somewhat more general context using
dynamical renormalization in \cite{Boyanovsky:2004gq}.

Similar studies \cite{Banks:2002nv,Einhorn:2002nu,Collins:2003zv} of
interacting theories in general $\alpha$-vacua, the de Sitter invariant
family, are plagued by non-renormalizable divergences that obscure any
attempt to probe their properties. Although this is a disturbing
feature it does not seem reasonable to simply discard them as they are
perfectly valid vacua of the free theory. There have been attempts to
address these divergences \cite{Einhorn:2003xb,Goldstein:2003ut} and
in particular we take up a suggestion put forward by Collins and
Holman in \cite{Collins:2003mj} when we study the problem of particle
decay. We find their prescription especially attractive as it yields a
vacuum independent result.

In this paper we will address the question of the slowly rolling
inflaton decaying into light scalars in a general $\alpha$-vacuum. We
will present a brief introduction to the nature of the vacuum
ambiguity of de Sitter space in section (\ref{BGsec}) and present the
notation we will employ. In section (\ref{SKsec}) we will present the
formalism necessary for calculating in a time dependent
background. The interaction lagrangian and the calculation of the
decay rate in the Bunch-Davies vacuum follows in section
(\ref{ILBDsec}). We then present the results in general $\alpha$-vacua
for conformally coupled and minimally coupled in sections
(\ref{CCsec}) and (\ref{MCsec}) respectively. Adopting a prescription
for addressing time ordering problems in the $\alpha$-vacua, we repeat
the calculation for the conformally coupled field in section
(\ref{TAsec}). Finally we end with some concluding remarks regarding a
consistent treatment of the $\alpha$-vacua and their significance.

\section{\label{BGsec}Background}
The quantization of a free scalar theory in a fixed de Sitter
background and the attendant vacuum choice ambiguity is described in
\cite{Mottola:1985ar,Allen:1985ux}. Here we sketch the procedure and
describe our conventions. Additional details of the procedure can also
be found in \cite{Bousso:2001mw,Collins:2003zv}.

We will find it convenient to work both in cosmic time(\ref{smet}) and
conformal time(\ref{nmet}) in which the metric is respectively given by
\begin{eqnarray}
  \label{smet}
  ds^2 & = & dt^2-e^{2Ht}d\vec{x}^2 \\
  \label{nmet}
       & = & \frac{d\eta^2-d\vec{x}^2}{H^2\eta^2}
\end{eqnarray}
where $H$, as usual, is the Hubble constant. The domains of the
coordinates are $t\in [-\infty,\infty]$ and $\eta\in [-\infty,0]$
and they are simply related by $\eta=-\frac{e^{-Ht}}{H}$.

A free massive scalar theory satisfies the Klein-Gordon equation
\begin{equation}
  \label{KGeqn}
  \left [\nabla^2+m_{\Phi}^2 \right ]\Phi=0.
\end{equation}
Since de Sitter space has constant curvature, any coupling to curvature
normally denoted by $\xi R$ can be absorbed into the mass and hence we
take $m_{\Phi}$ to denote an effective mass into which the coupling
has been absorbed. 

As usual the field is expanded in creation and annhilation operators
\begin{equation}
  \label{modeEx}
  \Phi(\eta,\vec{x})=\int\frac{d^3k}{(2\pi)^3}
  \left [U_k(\eta)e^{i\vec{k}\cdot\vec{x}}a_{\vec{k}}
        +U_k(\eta)^*e^{-i\vec{k}\cdot\vec{x}}{a_{\vec{k}}}^{\dag}\right ]
\end{equation}
where
\begin{equation}
  [a_{\vec{p}},{a_{\vec{q}}}^{\dag}]=(2\pi)^3\delta^3(\vec{p}-\vec{q}).
\end{equation}
Equation (\ref{KGeqn}) implies that the mode functions satisfy the
differential equation,
\begin{equation}
  \label{modeEq}
  \left [\eta^2\partial_\eta^2-2\eta\partial_\eta
         +\eta^2k^2+\frac{m_{\Phi}^2}{H^2}\right ]U_k(\eta)=0.
\end{equation}
This equation has Bessel functions as solutions. In particular, if we demand that the
modes match the Minkowski massive free scalar modes at short distances
then,
\begin{equation}
  U^E_k(\eta) = \frac{\sqrt{\pi}}{2}\eta^{3/2}H_{\nu}^{(2)}(k\eta),
\end{equation}
where,
\begin{equation}
  \nu = \sqrt{\frac{9}{4}-\frac{m^2}{H^2}}
\end{equation}
and $H_{\nu}^{(2)}$ denotes the Hankel funciton of the
second-kind. The normalization of the modes has been fixed by
enforcing the canonical commutation relation for the field and its
conjugate momentum which leads to a Wronskian condition for the mode
functions. We denote the corresponding annhilation operators
$a_{\vec{k}}^E$ and the vacuum annhilated by them, commonly referred
to as the Bunch-Davies vacuum, as $\ket{E}$. So far, quantization has
proceeded much as it does in Minkowski space but we have had to make
the reasonable but arbitrary assumption that the modes match Mikowski
modes at short distances to fix the vacuum.

\subsection{Alpha vacua}

When we relax the Minkowski matching condition we see that any norm
preserving linear combination of the mode function and its conjugate
will satisfy (\ref{modeEq}) and preserve the commutation
relations. The standard literature addresses this by considering a
Bogolubov transformation of the creation-annhilation operators,
\begin{equation}
  a_{\vec{k}}^{\alpha} = N_{\alpha}
  \left( a_{\vec{k}}^E-e^{\alpha^*}{a_{-\vec{k}}^E}^{\dag} \right),
\end{equation}
where $\Re[\alpha]<0$. The vacuum annhilated by all
$a_{\vec{k}}^{\alpha}$ is denoted $\ket{\alpha}$. Rewriting
(\ref{modeEx}) in terms of $a_{\vec{k}}^{\alpha}$ we find the
corresponding mode functions,
\begin{equation}
  U^{\alpha}_k(\eta)=N_\alpha \left (U^E_k(\eta)
                    +e^{\alpha} {U^E_k(\eta)}^* \right ).
\end{equation}

It remains to be shown that this set of vacua parametrized by $\alpha$
-- the Bunch-Davies vacuum is recovered at the $\alpha \rightarrow
-\infty$ limit -- are infact de Sitter invariant. This can be
demonstrated by showing that the two-point functions can be written in terms of the  de Sitter
invariant distance between the two points. We will construct the two point functions in the next
section and address the question briefly while details can be found in
\cite{Collins:2003zv} and references therein.

\subsection{\label{GFsection}Green's functions}

To compute the diagrams we will eventually encounter in evaluating the
decay rate we need to construct the propagators for the free
theory. Since the metric we are considering is spatially flat we can
write
\begin{equation}
  \mathcal{G}^{\alpha}(x,x')=\bra{\alpha}\Phi(x)\Phi(x')\ket{\alpha}
    =\int\frac{d^3k}{(2\pi)^3}e^{i\vec{k}\cdot(\vec{x}-\vec{x'})}
                             U^{\alpha}_k(\eta)U^{\alpha}_k(\eta')^*.
\end{equation}
This function depends only on $Z(x,x')$, the de Sitter invariant
distance, and hence the de Sitter invariance of the $\alpha$-vacua. In a momentum representation we have
\begin{equation}
  \label{GFrep}
  \mathcal{G}^{\alpha}_k(\eta,\eta')
      =U^{\alpha}_k(\eta)U^{\alpha}_k(\eta')^*,
\end{equation}
as the momentum space Wightman function.

From the Wightman function we can construct the Feynman propagator
following the same procedure as in Minkowski space,
\begin{eqnarray}
  \label{stdTO}
  G^{\alpha}(x,x')&=&i\bra{\alpha}T[\Phi(x)\Phi(x')]\ket{\alpha}, \\
  &=&\int\frac{d^3k}{(2\pi)^3}e^{i\vec{k}\cdot(\vec{x}-\vec{x'})}
         G^{\alpha}_k(\eta,\eta'), \nonumber \\
  G^{\alpha}_k(\eta,\eta')&=&\Theta(\eta-\eta')\,\mathcal{G}(\eta,\eta')
                           + \Theta(\eta'-\eta)\,\mathcal{G}(\eta',\eta) .
\end{eqnarray}
Defined in this manner the Feynman propagator, as would be expected,
satisfies the equation of motion with a point source,
\begin{equation}
\left [ \nabla_x^2+m_\Phi^2 \right ]G^{\alpha}(x,x') = 
   \frac{\delta^{(4)}(x-x')}{\sqrt{-g(x)}}.
\end{equation}

Although natural, this definition for the Feynman propagator is not
the only one leading to a well-defined object, and further, it is
perhaps not the most meaningful defintion due to subtleties associated
with the $\alpha$-vacua and causality. These issues are addressed in
\cite{Einhorn:2003xb,Goldstein:2003qf,Collins:2003mj} and an $\alpha$
dependent time ordering in the definition of the propagator is
suggested. Here we will present the central idea and the details
necessary for calculation.

In de Sitter space there is a natural antipode map given in the
conformal time coordinatization by $x_A:(\eta,\vec{x})
\rightarrow (-\eta,\vec{x})$. We can then expand
\begin{eqnarray}
\nonumber
\mathcal{G}^{\alpha}(x,x') & = &
N_{\alpha}\bigl(\mathcal{G}_E(x,x')+e^{\alpha+\alpha^*}G_E(x',x) \\
& & + e^{\alpha}\mathcal{G}_E(x_A,x')
    + e^{\alpha^*}\mathcal{G}_E(x',x_A)\bigr).
\end{eqnarray}
This result indicates that ordinary time-ordering will not impose the
causality expected. Instead we can define a time-ordering prescription
that ensures a causal definition for the Feynman propagator and it
leads to a double source propagator,
\begin{equation}
\left [ \nabla_x^2+m_\Phi^2 \right ]G^{\alpha}(x,x') = 
   A_{\alpha} \frac{\delta^{(4)}(x-x')}{\sqrt{-g(x)}}
  +B_{\alpha} \frac{\delta^{(4)}(x_A-x')}{\sqrt{-g(x_A)}},
\end{equation}
where $A_{\alpha}$ and $B_{\alpha}$ must be fixed by some prescription
while maintaining
\begin{equation}
  \lim_{\alpha\rightarrow -\infty}A_{\alpha}=1
  \quad\text{and}\quad
  \lim_{\alpha\rightarrow -\infty}B_{\alpha}=0
\end{equation}
so that the standard time-ordering is recovered in this limit. Doubly
sourced propagators have been discussed in \cite{Banks:2002nv}, but
without the modified time ordering they lead to non-renormalizable
divergences. We choose the particularly simple scheme $A_{\alpha}=1$
and $B_{\alpha}=0$. This leads to a singly sourced propagator and in the
path integral formulation all internal lines are naturally converted
to the Bunch-Davies propagator. This conveniently takes care of the
divergences that plague calculations in the general $\alpha$-vacua. We
should note however that the two-point functions are still $\alpha$
dependent.

\subsection{\label{SKsec}Schwinger-Keldysh formalism}

Since de Sitter space presents a time dependent background it is not
meaningful to calculate S-matrix elements. The absence of a global
time like Killing vector does not allow us to relate aymptotic
\emph{in} and \emph{out} and states and at any rate we mean to
calculate the matrix element between time evolving states at a given
time. The Schwinger-Keldysh \cite{Schwinger:1961qe,Keldysh:1964ud,
Mahanthappa:1962,Bakshi:1963bn} formalism addresses these issues by
calculating the evolution of matrix elements from given initial
conditions to some later time. The application of the formalism to
de Sitter space is discussed in \cite{Boyanovsky:1997ab,Collins:2003zv}.

\begin{figure}
\centering
\includegraphics[width=3in]{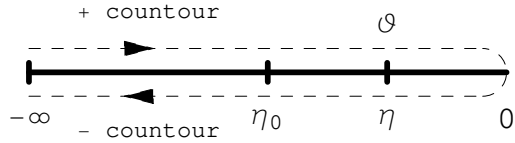}
\caption{The closed time contour that appears in the evolution of 
operators over a finite time interval. Separate copies of the field
appear on the forward running and backward running portions of the
contour. Here the contour is labelled in conformal coordinates.
\label{ctp}}
\end{figure}

We can summarize the results by considering the evolution of the
density matrix. If we define the evolution operator so that
\begin{equation}
  i\frac{\partial}{\partial\eta}U_I(\eta,\eta_0)=H_I U_I(\eta,\eta_0),
\end{equation}
and we take the full Hamiltonian to be given by
\begin{equation}
  H=H_0+\Theta(\eta-\eta_0)H_I.
\end{equation}
We define initial conditions at $\eta_0$ where $\rho(\eta_0)=\rho_0$.
Then we can calculate the time-dependent expectation value of a
general operator,
\begin{equation}
  \expect{\mathcal{O}}(\eta) =
  \frac{\text{Tr}\bigl [U_I(-\infty,0)U_I(0,\eta)
                        \mathcal{O}U_I(\eta,-\infty)\rho_0 \bigr]
       }{\text[Tr]\bigl [U_I(-\infty,0)U_I(0,-\infty)\rho_0 \bigr]}
\end{equation}
The numerator of this expression represents a closed time contour
(fig. \ref{ctp}) running from the infinite past with given initial
conditions, enclosing the operator inserted at $\eta$, the infinite
future and finally running back to the infinite past.

The calculation is implemented formally by considering fields that
live on the forward contour and a separate set of fields that live on
the backwards contour, labelled respectively by '$+$' and '$-$'. Time
ordering is then implemented so that '$-$' fields always appear after
'$+$' fields and that time runs backward on the backward contour. The
respective Green's functions are given by,
\begin{eqnarray}
  G^{(++)}_k(\eta,\eta') & = &
  \Theta(\eta-\eta')\mathcal{G}_k(\eta,\eta')+
  \Theta(\eta'-\eta)\mathcal{G}_k(\eta',\eta) \\
  G^{(--)}_k(\eta,\eta') & = &
  \Theta(\eta'-\eta)\mathcal{G}_k(\eta,\eta')+
  \Theta(\eta-\eta')\mathcal{G}_k(\eta',\eta) \\
  G^{(-+)}_k(\eta,\eta') & = & \mathcal{G}_k(\eta,\eta') \\
  G^{(+-)}_k(\eta,\eta') & = & \mathcal{G}_k(\eta',\eta)
\end{eqnarray}
where we notice that $G^{++}$ is the usual Feynman propagator.
Further, due to the doubling of fields, we have '$+$' and '$-$'
vertices corresponding to each of the vertices in the original
interaction lagrangian. Further details of this formal procedure will
become apparent when we present the calculation of the decay rate.

\section{\label{ILBDsec}Interaction Lagrangian}
We now describe the model we will use to study inflaton decay
\cite{Boyanovsky:1997ab} which consists of the inflaton field coupled
to a massless scalar field, $\sigma$ via a three point
interaction. We anticipate the requirements of the Schwinger-Keldysh
formalism and include both '$+$' and '$-$' fields in the lagrangian,
\begin{eqnarray}
  \nonumber
  \mathcal{L} & = & 
  \frac{1}{2}\bigl(\partial_{\mu}\Phi^+\partial^{\mu}\Phi^+
                   +m_{\Phi}^2{\Phi^+}^2
		  +\partial_{\mu}\sigma^+\partial^{\mu}\sigma^+
                   +\xi_{\sigma}R{\sigma^+}^2 \bigr) \\
  & &+\frac{1}{2}g\Phi^+{\sigma^+}^2+h(t)\Phi^\\
  \nonumber & &-\bigl [+ \rightarrow - \bigr ]
\end{eqnarray}  
where the negative sign appearing before $\mathcal{L}(\Phi^-,\sigma^-)$
is a result of the time reversal on the negatively directed time contour,
\begin{equation}
  \begin{split}
    S_I&=-\int_{-\infty}^0H_I(\Phi^+,\sigma^+)d\eta
         -\int_0^{-\infty}H_I(\Phi^-,\sigma^-)d\eta\\
       &=-\int_{-\infty}^0\bigl [H_I(\Phi^+,\sigma^+)
                                -H_I(\Phi^-,\sigma^-)\bigr]d\eta\, .
  \end{split}
\end{equation}
The interaction term is apparent, but the magnetic term involving
$h(t)$ requires some explanation. Since we are primarily interested in
the decay of $\Phi$ into $\sigma$ we need not be concerned with the
details of the potential for $\Phi$ as long as the field is rolling
slowly with respect to the decay rate into $\sigma$. So instead of
including an explicit, but arbitrary, potential we simply include a
magnetic (or source) term to set initial conditions such that $\Phi$
is 'slowly rolling'. In addition $h(t)\Phi$ term is required for the
renormalization of tadpole diagrams\cite{Boyanovsky:1997ab}.
 
As usual the inflaton field is separated into a homogenous term and
scalar fluctuations,
\begin{equation}
  \Phi^{\pm}(\vec{x},t) = \phi(t)+\varphi^{\pm}(\vec{x},t)
  \quad\text{where}\quad
  \expect{\Phi^{\pm}} = \phi(t).
\end{equation}
This implies the tadpole condition 
\begin{equation}
  \expect{\varphi^{\pm}(\vec{x},t)}=0
\end{equation}
from which we obtain the equation of motion. In addtion we also
enforce $\expect{\sigma(\vec{x},t)}=0$ which ensures that the
$\sigma$ field does not acquire an expectation value.

Rewriting the action in terms of the shifted fields we find
\begin{equation}
  \begin{split}
    S=\int d^4x\sqrt{-g}\Bigl\{&
        \mathcal{L}_0(\varphi^+)+\mathcal{L}_0(\sigma^+) \\
      &+\varphi^+\bigl [-\ddot{\phi}-3H\dot{\phi}-m_{\Phi}^2\phi \bigr]\\
      &+\frac{g}{2}(\phi+\varphi^+)(\sigma^+)^2+h\varphi^+\\
      &-[+ \rightarrow -]\Bigr\},
  \end{split}
\end{equation}
where $\mathcal{L}_0$ denotes the lagrangian of the corresponding free
theories. Now evaluating the tadpole condition
\begin{equation}
\begin{split}
  \expect{\varphi^+(\vec{x},\tau)}=&
  \int\mathcal{D}\varphi^{\pm}\mathcal{D}\sigma^{\pm}
                     \varphi^+(\vec{x},\tau)e^{iS}\\
  =&\int d^3ydt\sqrt{-g}
    i\Bigl [\ddot{\phi}+3H\dot{\phi}+m_{\Phi}^2\phi
      -\frac{g}{2}\expect{\sigma^2}(t)-h(t)\\
  &\quad\quad-i\frac{g^2}{4\pi^2}\int_{-\infty}^t
                              dt'\sqrt{-g}K(t,t')\phi(t')
     \Bigr ]\times\\
  &\quad\bigl(\expect{\varphi^+(\vec{x},\tau)\varphi^+(\vec{y},t)}
             -\expect{\varphi^+(\vec{x},\tau)\varphi^-(\vec{y},t)})\\
  &\quad+\mathcal{O}(g^3)	     
\end{split}
\end{equation}
where we have defined
\begin{equation}
  K(t,t')\equiv\int_0^{\infty}dkk^2\left(G_k(t',t)^2-G_k(t,t')^2\right).
\end{equation}
To enforce the tadpole condition we must require that the coefficient
of $\bigl(\expect{\varphi^+(\vec{x},\tau)\varphi^+(\vec{y},t)}
-\expect{\varphi^+(\vec{x},\tau)\varphi^-(\vec{y},t)})$ vanish and
this yields the equation of motion for $\phi(t)$. First we must take
care of the $\expect{\sigma^2}$ term which we simply absorb into
$h(t)$ defining
\begin{equation}
  h_{\sigma}(t)=h(t)+\frac{g}{2}\expect{\sigma^2}(t).
\end{equation}
The inhomogenous term does not enter the expression for the decay rate
and so is not of concern here. Its main function is to act as a
Lagrange multiplier to enforce the initial conditions at $t=0$,
\begin{eqnarray}
  \phi(0)&=&\phi_i\\
  \dot{\phi}(0)&=&0\\
\end{eqnarray}

It turns out that the kernel $K(t,t')$ can be cast in
convolution form $K(t-t')$ and so the equation can be solved by
Laplace transforming $\phi(t)$. In general we define
\begin{equation}
  \tilde{f}(s)=\mathcal{L}\{f(t)\}\equiv\int_0^{\infty}dte^{-st}f(t).
\end{equation}
The exact form of the kernel depends on the nature of the coupling of
the $\sigma$ field to curvature and its mass. However, generally the
kernel integral possesses a logarithmic UV divergence which can be
absorbed into a mass redefinition. In the Bunch-Davies vacuum the
renormalized mass is given by
\begin{equation}
  m_{\Phi,R}^2=m_{\Phi}^2+\frac{g^2}{16\pi^2}(\ln\epsilon+1)
\end{equation}
The details of Laplace transforming the equation of motion also depend
on properties of the $\sigma$ field but as an illustrative example for
the conformally coupled massless case in the Bunch-Davies vacuum
\cite{Boyanovsky:1997ab} it is found that
\begin{eqnarray}
  &&s^2\tilde{\phi}(s)-s\phi_i+3H(s\tilde{\phi}(s)-\phi_i)
    +m_{\Phi,R}^2\tilde{\phi}(s)
    +\frac{g^2}{16\pi^2}\Sigma(s)\tilde{\phi}(s)=\tilde{h_{\sigma}}(s)\\
  &&\Sigma(s)=-s\mathcal{L}\{\ln(1-e^{-Ht})+e^{-Ht}\}.
\end{eqnarray}

Solving this algebraic equation for $\tilde{\phi}(s)$ and taking the
inverse Laplace transform the decay rate can be identified from
resulting solution for $\phi(t)$. In the particular example we are
discussing the decay rate is found to be\cite{Boyanovsky:1997ab}
\begin{equation}
  \Gamma=\frac{g^2}{16\pi^2}\frac{\Im[\Sigma(s_0^+)]}{M_{\Phi}}
        =\frac{g^2}{32\pi M_{\Phi}}
           \tanh\left(\frac{\beta_H M_{\Phi}}{2}\right)
\end{equation}
where $\beta_H=\frac{2\pi}{H}$ and to this order
\begin{eqnarray}
  M_{\Phi}^2 & = & m_{\Phi}^2-\frac{9H^2}{4} \\
  s_0^+ & = & -\frac{3H}{2}+iM_{\Phi}.
\end{eqnarray}

The same general procedure is employed in the following
sections. However, the mode functions and hence the Green's functions
vary for different theories (conformally coupled/minimally coupled)
and $\alpha$-vacua. This implies that the kernel acquires new features and, in the
case of a minimally coupled field, infrared divergences. However,
after some work the decay rate can be extracted in the same fashion
from the solution to the equation of motion. These results are
pressented in the next two sections.

\subsection{\label{CCsec}Conformally coupled massless $\sigma$}
Specializing to the conformally coupled massless case, the calculation
for general $\alpha$ proceeds in the same manner as detailed above for
the Bunch-Davies vacuum, except the appropriate $\alpha$ dependent
Green's functions are employed in computing the diagram. Defining
\begin{eqnarray}
  E_{\alpha} & = & \frac{1+e^{\alpha+\alpha^*}}{1-e^{\alpha+\alpha^*}} \\
  D_{\alpha} & = & \frac{e^{\alpha}+e^{\alpha^*}}{1+e^{\alpha+\alpha^*}}
\end{eqnarray}
we can write the $\alpha$ dependent mass renormalization and
self-energy as
\begin{eqnarray}
  m_{\Phi,R}^2 & = & m_{\Phi}^2+E_{\alpha}\frac{g^2}{16\pi^2}(\ln\epsilon+1) \\
  \Sigma(s) & = & E_{\alpha}\left(\Sigma_E(s)+D_{\alpha}\tilde{\chi}(s)\right)
\end{eqnarray}
where
\begin{equation}
  \tilde{\chi}(s) = \frac{H}{s+2H}-\frac{H}{s+H}
\end{equation}
is a non-multiplicative correction to the self-energy. It can be seen
that the mass renormalization differs from the euclidean case only
multiplicatively and it should be noted that as
$\alpha\rightarrow-\infty$ (the Bunch-Davies limit) the multiplicative
factor $E_{\alpha}\rightarrow 1$. Although the modification to the
self-energy is not purely multiplicative in general, the
imaginary part {\em is} only multiplicatively modified.
\begin{equation}
\Gamma = E_{\alpha}\frac{g^2}{32\pi m_{\Phi}}
             \tanh\left(\frac{\beta_H M_{\Phi}}{2}\right)
       = E_{\alpha} \Gamma_E
\end{equation}
Rewriting the multiplicative factor in a suggestive form
\begin{equation}
\label{BEfactor}
E_{\alpha} = 1+2(e^{-(\alpha+\alpha^*)}-1)^{-1}
\end{equation}
we can interpret the decay rate as being Bose enhanced with respect to
the Bunch-Davies rate. It should be noted that the Bunch-Davies rate
is unique in that it is the minimum rate since $E_{\alpha}\ge 1$.

\subsection{\label{MCsec}Minimally coupled massless $\sigma$}

The mode solution for the minimally coupled scalar in the Bunch-Davies
vacuum is given by
\begin{equation}
  U^E_k(\eta)=(i-k\eta)\frac{e^{-ik\eta}}{\sqrt{2k^{3/2}}}
\end{equation}
and from this we can construct the general mode functions
$U^{\alpha}_k(\eta)$. Then the Green's functions follow and once again
the procedure as outlined above can be invoked. However, new
complications arise, the first of which is an infrared divergence
presumably due to the massless minimal coupling. The divergence can be
regulated by imposing a lower cut-off on the loop integral. Further,
the resulting equation of motion after Laplace transforming is no
longer algebraic but rather a first order differential equation of the
form,
\begin{equation}
  \left [p_0(s)+\lambda
    \left (p_1(s)+q(s)\frac{d}{ds}\right )\right ]\tilde{\phi}(s)=f(s)
\end{equation}
Solving this equation, order by order in $\lambda$, 
\begin{eqnarray}
  \tilde{\phi}(s) & = & \tilde{\phi}_0(s)
                    +\lambda\tilde{\phi}_1(s)
                    +\mathcal{O}(\lambda^2)\\
  \tilde{\phi}_0(s) & = & \frac{f(s)}{p_0(s)} \\
  \tilde{\phi}_1(s) & = & -\frac{\tilde{\phi}_0(s)}{p_0(s)}
       \left [p_1(s)+q(s)\frac{d\ln\tilde{\phi}_0}{ds} \right ]
\end{eqnarray}  
leads to the result
\begin{equation}
  \tilde{\phi}(s)=\frac{f(s)}{p_0(s)+\lambda
    \left [ p_1(s)+q(s)\frac{d}{ds}\ln\frac{f(s)}{p_0(s)}\right ]}
  +\mathcal{O}(\lambda^2),
\end{equation}
This is of the same form as the conformally coupled case, at least to
$\mathcal{O}(g^4)$. The decay rate, once again, acquires a purely
multiplicatively enhancement, and we find,
\begin{equation}
  \Gamma = E_{\alpha}\frac{g^2}{32\pi M_{\Phi}}
     \left [\left(1+\frac{4H^2}{m_{\Phi}^2}\right)
                \tanh\left(\frac{\beta_H M_{\Phi}}{2}\right)
          +\frac{8H^3 M_{\Phi}}{\pi m_{\Phi}^4}\right]
\end{equation}

\subsection{\label{TAsec}Alpha Time Ordering}

The abscence of divergences in the preceding calculation of the decay
rate in an alpha vacuum is unusual considering that previous
investigations \cite{Danielsson:2002mb,Einhorn:2002nu,Collins:2003zv}
seem to indicate that non-renormalizable divergences inevitably appear
in alpha vacuum calculations. The significance of these divergences is
still unclear but in certain cases they can be traced to pinched
singularities in the propagator.

As discussed in section (\ref{GFsection}) it is possible to define an
$\alpha$-dependent time ordering prescription that resolves the
pinched singularities appearing in the Feynman propagator. A related
procedure is discussed in \cite{Goldstein:2003ut} where non-local
interactions are introduced to absorb the divergences that arise while
preserving the locality of the theory. In fact, the inclusion of a
non-local (point/anti-pode) interactions that preserve the locality of
the theory in the sense that commutators vanish outside the light cone
was discussed in \cite{Goldstein:2003qf,Collins:2003mj}. So for this
section we will generalize the interaction lagrangian to include these
antipodal interactions.

Before we present the new interaction lagrangian, it is important to
note that the main result of the time ordering prescription together
with our choice for the arbitrary constants $A_{\alpha}=1$ and
$B_{\alpha}=0$ is that internal propagators are converted to the
Bunch-Davies propagator and all the $\alpha$-dependence is restricted
to the external legs. This is distinct from a field redefinition in
that we hold the interaction term fixed when we change the ordering
prescription. A field redefinition would demand the interaction term
undergo the corresponding transformation. This point is discussed in
\cite{Goldstein:2003qf} where it is clarified that a redefinition of
the kinetic term to make it once again appear local would lead to a
non-local interaction term. The appearance of euclidean propagators on
internal lines indicates that the only divergences that will appear
are the ones encountered in the Bunch-Davies vacuum which can be
renormalized via the usual mass and (if relevant) wave function and
coupling constant renornalization.

We write the new interaction lagrangian that includes antipodal
interactions as
\[ {\cal L}_I = \frac{1}{2}e^{3Ht}g\,\sum_{i,j\in\{{\cal P,\,A}\}}\lambda_{ij}\Phi(x)\,\sigma(x_i)\sigma(x_j) \]
where we take $x_{\cal P}=(\eta,\vec{x})$ and $x_{\cal A}=(-\eta,\vec{x})$
and $\lambda_{ij}$ is a dimensionless matrix describing the relative 
strengths of the interaction terms. To fix the spurious freedom 
in the definition of $\lambda$ we impose
\begin{eqnarray}
  \lambda_{\mathcal{PP}} & = & 1 \\
  \lambda & = & \lambda^T
\end{eqnarray}
This particular choice of conditions ensures that the earlier result
is recovered when all the free components of $\lambda$ approach zero.

Now there are a sum of diagrams to be evaluated resulting in the 
kernel given by
\[ 
\begin{aligned}
  K(\eta,\eta')&= \sum_{ii'jj'}\int_0^{\infty}k^2dk\lambda_{ij}\lambda_{i'j'} \\
    &\quad\bigl\{G^{++}_k(\eta_i,\eta'_{i'})G^{++}_k(\eta_j,\eta'_{j'})
                -G^{+-}_k(\eta_i,\eta'_{i'})G^{+-}_k(\eta_j,\eta'_{j'})\bigr\}
\end{aligned}
\]
For the massless conformally coupled case, $\nu=\frac{1}{2}$, only UV divergences appear in the 
integral which can be regulated as was done
previously with $e^{ika} \rightarrow e^{ik(a+i\epsilon)}$. In this
case we find
\begin{equation}
  \begin{aligned}
    \frac{1}{\eta^2\eta'^2} & K(\eta,\eta') = \\
      & \lambda_\mathcal{PP}^2 \Theta(\eta-\eta')\frac{4i(\eta-\eta')}{\epsilon^2+4(\eta-\eta')^2} \\
    - & \lambda_\mathcal{PP}\lambda_\mathcal{AA} \frac{4i(\eta+\eta')}{\epsilon^2+4(\eta+\eta')^2} \\
    + & \lambda_\mathcal{AA}^2 \Theta(\eta'-\eta)\frac{4i(\eta'-\eta)}{\epsilon^2+4(\eta'-\eta)^2} \\
    + 4\lambda_\mathcal{PA} \Bigl\{ &
      \lambda_\mathcal{PP}\bigl[\frac{1}{2}\bigl(\frac{1}{\epsilon-2i\eta}-\frac{1}{\epsilon+2i\eta'}\bigr)
	             -\Theta(\eta-\eta')\bigl(\frac{1}{\epsilon-2i\eta}-\frac{1}{\epsilon-2i\eta'}\bigr)\bigr]\\
    +&\lambda_\mathcal{AA}\bigl[\frac{1}{2}\bigl(\frac{1}{\epsilon-2i\eta'}-\frac{1}{\epsilon+2i\eta}\bigr)
	             -\Theta(\eta-\eta')\bigl(\frac{1}{\epsilon-2i\eta'}-\frac{1}{\epsilon-2i\eta}\bigr)\bigr]\\
    +&\lambda_\mathcal{PA}\bigl[\frac{1}{\epsilon}+\Theta(\eta-\eta')\frac{4i(\eta-\eta')}{\epsilon^2+4(\eta-\eta')^2}\\
                      &\quad-\frac{1}{2}\bigl(\frac{1}{\epsilon-2i(\eta+\eta')}
                                             +\frac{1}{\epsilon-2i(\eta-\eta')}\bigr)\bigr]\Bigr\}
  \end{aligned}
\end{equation}

First of all we should note that the non-local interactions mixing
point/antipode fields -- the $\lambda_\mathcal{PA}$ terms -- give rise to a
linear divergence in the kernel. More problematically, all terms
involving the antipodal field give rise to terms of the form
\[ \int_\eta^0 d\eta' \phi(\eta')f(\eta',\eta) \]
in the equation of motion which manifestly violate causality, atleast
in our framework with adiabatic switching in the infinite
past. Boundary conditions of this form have been considered in the
context of elliptic de Sitter space, see \cite{Parikh:2002py} and
references therein. Although the theory is local in the sense that
communators vanish outside the light cone, we are led to conclude that
the resulting theory is not causal in our framework since we cannot
define a useful initial value boundary condition. So it appears that
the only meaningful choice for the interaction matrix is
\[\lambda_\mathcal{AA} = \lambda_\mathcal{PA} = 0.\]

With this choice for the interaction the prescription for modifying
the propagators recovers the euclidean result for the decay
rate. Further, in \cite{Collins:2003mj} it was noted that euclidean
self energy is also recovered. This, of course, is not surprising
considering that in essence the prescription sets all to internal
propagators to the euclidean propagator. This suggests that proper
attention to causality -- which motivated the propagator redefinition
-- removes any vacuum ambiguity in an interacting theory. It provides
an additional motivation for calculating in the euclidean vacuum,
namely, it is the vacuum that yields the natural feynman propagator
since its definition depends of enforcing causality. However, we can
still choose to calculate in any vacuum but physical results will be
independent of our choice.

\section{Conclusions}

In this paper we have considered the decay of the inflaton into light
scalars placed in a general de Sitter invariant $\alpha$-vacuum.  The
result for the commonly considered Bunch-Davies vacuum,
\cite{Boyanovsky:1997ab}, yields a decay rate that corresponds to
decay in Minkowski space at a finite temperature, $\beta_H=2\pi/H$.
For a general $\alpha$-vacuum we find a further enhancement that is
analogous to Bose enhancement seen in stimulated emission with the
enhancement factor given in eq.(\ref{BEfactor}). We find that this is the
case for both conformally and minimally coupled scalars. This
interpretation is compelling considering that when $\alpha$-vacua are
projected on the Hilbert space constructed from the Bunch-Davies
vacuum they correpond to states with constant occupation of all modes
which would lead to stimulated decay. Of course, this still singles
out the Bunch-Davies vacuum as it is relative to the Bunch-Davies rate
that the rate for general $\alpha$ are enhanced.

We have also considered a recently proposed prescription for treating
the $\alpha$-vacua that seeks to address the proper time-ordering of
the Feynman propagator in a general $\alpha$-vacuum. The prescription
leads to the internal lines of diagrams being converted to the
Bunch-Davies propagator independent of $\alpha$. This leads to a
self-energy, caclculated in \cite{Collins:2003mj}, and decay rate,
as discussed in this paper, that are independent of $\alpha$. Further,
the only divergences that can arise from loops are those that are
encountered in the Bunch-Davies vacuum and hence the theory is
renormalizable for all $\alpha$. This is an appealing result as it
leads to an $\alpha$ independent scheme for calculation. However,
there is still $\alpha$ dependence in the external legs which will
appear for example in the two-point function relevant to
inflation. This seems to be related to the manner of initial
conditions and is question that we are still investigating.



%





\bibliography{../bib/inflation}

\end{document}